\documentclass[reprint,aps,pra]{revtex4-1}

\usepackage{graphicx}
\usepackage{epsfig}
\usepackage{epstopdf}
\usepackage{dcolumn}
\usepackage{bm}
\usepackage{lineno}
\usepackage{amsmath, amssymb}
\usepackage{accents}
\usepackage{lipsum}
\usepackage{multirow}
\usepackage[usestackEOL]{stackengine}
\usepackage{tabularx}
\usepackage{makecell}
\usepackage{xcolor}
\usepackage{dcolumn}
\usepackage{bm}
\usepackage{booktabs}
\usepackage{comment}
\usepackage{arydshln}
\usepackage{yhmath}
\usepackage{float}
\usepackage[super]{nth}
\usepackage{subcaption}
\usepackage{hyperref}
\usepackage{chemfig}
\usepackage{physics}
\usepackage{caption}

\begin{document}
\title{Quantum theory of a potential biological magnetic field sensor: radical pair mechanism in flavin adenine dinucleotide biradicals}
\author{Amirhosein Sotoodehfar$^{1, 2, 3, *}$}
\author{Rishabh$^{1, 2, 3, *}$}
\author{Hadi Zadeh-Haghighi$^{1, 2, 3, *}$}
\author{Christoph Simon$^{1, 2, 3, *}$}
\affiliation{$^1$Department of Physics and Astronomy, University of Calgary, Calgary, AB, T2N 1N4, Canada}
\affiliation{$^2$Institute for Quantum Science and Technology, University of Calgary, Calgary, AB, T2N 1N4, Canada}
\affiliation{$^3$Hotchkiss Brain Institute, University of Calgary, Calgary, AB, T2N 1N4, Canada}
\affiliation{$^*$\rm{amirhossein.sotoodeh@ucalgary.ca} $\&$ \rm{rishabh1@ucalgary.ca} $\&$ 
\rm{hadi.zadehhaghighi@ucalgary.ca} $\&$ \rm{csimo@ucalgary.ca}
}

\begin{abstract}
\textbf{Abstract}  Recent studies \textit{in vitro} and \textit{in vivo} suggest that flavin adenine dinucleotide (FAD) on its own might be able to act as a biological magnetic field sensor. Motivated by these observations, in this study, we develop a detailed quantum theoretical model for the radical pair mechanism (RPM) for the flavin adenine biradical within the FAD molecule. We perform molecular dynamics simulations to determine the distance between the radicals on FAD, which we then feed into a quantum master equation treatment of the RPM. In contrast to previous semi-classical models which are limited to the low-field and high-field cases, our quantum model can predict the full magnetic field dependence of the transient absorption signal. Our model's predictions are consistent with experiments.
\end{abstract}
\maketitle 

\section{Introduction} 
Magnetosensitivity is abundant throughout biology and many biological systems are under the influence of Earth's weak magnetic field in various aspects, using it as a sensory cue for migration \cite{ref:senscue1, ref:senscue2, ref:reviewsimon, ref:clarice} to the regulation of plant function and growth \cite{ref:mfplant1, ref:mfplant2}. 
It is known that different animals such as migratory birds \cite{ref:birds1, ref:birds2}, sea turtles \cite{ref:seaturtules1, ref:seaturtles2}, and some insects \cite{ref:insects1, ref:insects2} sense and use Earth's magnetic field. 

A number of models have been proposed to explain magnetoreception in biological systems. The most prominent among them is magnetoreception based on radical-pair mechanism (RPM) \cite{ref:mainmodels1, ref:birds1, ref:hore_RPM, ref:Betony1} initially proposed by Schulten et al. \cite{ref:shculten1978}. 
Spin-correlated pairs of radicals (molecules with an unpaired electron) can be created via electron transfer 
from one closed shell molecule to another or homolytic cleavage of a chemical bond. 

Comparing the thermal energy at room temperature, $10^{-20}$J, and the magnetic interaction energy, $10^{-27}$J, may lead one to expect that Earth's magnetic field should have a negligible impact on biology and chemistry (or biochemistry). However, as we see in Sec. \ref{sec:theory}, RPM model explains how Earth's weak magnetic field can change the relative yield of chemical products for certain reactions.

Thus far, the primary candidate for RPM magnetoreception is a flavoprotein molecule known as Cryptochrome in which a blue light activated electron transfer between flavin adenine dinucleotide (FAD) molecule and tryptophan triad leads to the formation of radical pair. \cite{ref:mainmodels4, ref:hore_RPM, ref:ritz, ref:Nature2021, ref:Katting1}.

Recently, an experimental work using electrophysiology and behavioral analyses, conducted by Bradlaugh et al. challenges this Cryptochrome-based RPM model for magnetoreception \cite{ref:essential_biradical}. Their results indicate that FAD alone can also act as a biological magnetosensor.

Bradlaugh et al.'s observations suggest a possibility of a potential FAD-based RPM. The formation of radical pairs can be achieved by intramolecular electron transfer from adenine to flavin moiety in FAD molecule under blue light excitation in aqueous solution \cite{ref:woodward}. Previously, it has been shown that the photochemistry of FAD is sensitive to external magnetic field \cite{ref:woodward, ref:maeda2005}. The RPM model enables us to understand the role of the external magnetic field in the photochemistry of FAD. The external magnetic field alters the interconversion between singlet/triplet states. This interconversion between spin states is under the influence of several interactions, such as Zeeman, hyperfine, exchange, and dipole-dipole couplings. 

The previous theoretical models for the photochemistry of FAD are based on a semi-classical approach using rate equations \cite{ref:maeda2005}. These semi-classical models can only describe two extreme cases of magnetic field dependence (high field and low field). Also, in these models, the hyperfine interaction and relaxation process are treated phenomenologically via rate equations. Furthermore, they cannot incorporate the effect of other important interactions such as exchange and dipole-dipole coupling.

\begin{figure}[H]
    \centering
    \includegraphics[width=0.4\textwidth]{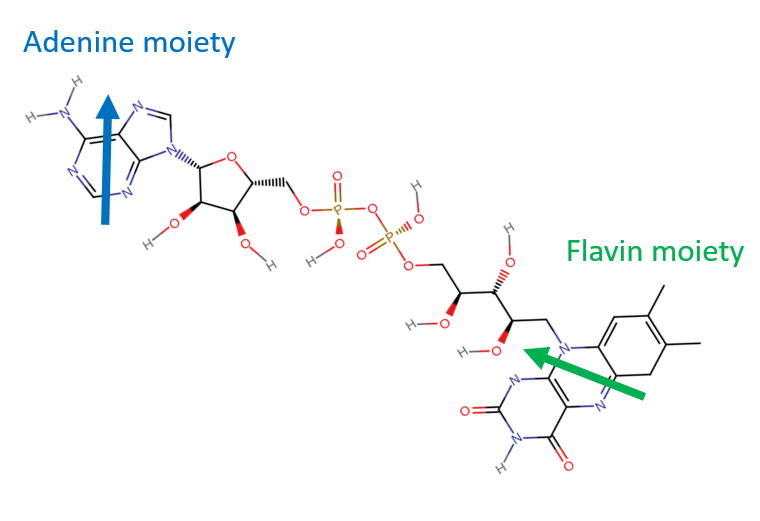}
    \caption{Structure of flavin adenine dinucleotide. This structure is made by using the Ligand Reader and Modeler available in CHARMM-GUI \cite{ref:charmmgui1, ref:charmmgui2}. The approximate location of the radicals in the flavin and adenine moieties are depicted by arrows.}
    \label{fig:FAD}
\end{figure}

As the exchange and dipole-dipole interactions depend on the distance between two radicals, an understanding of the conformation of FAD molecule is needed. We have performed molecular dynamics (MD) calculations to predict the movement of FAD molecule in solvents. This helps us to understand the distance between the radical spins within the molecule as a function of time. The structure of FAD is shown in Fig. \ref{fig:FAD}. By knowing this distance, we have the functionality of exchange and dipole-dipole interactions in time, which we can then integrate into our quantum RPM model. For this RPM model, we used a quantum master equation to obtain the spin dynamics of the radical pair system under the influence of the above-mentioned interactions. Notice that time-dependent behavior of exchange and dipole-dipole interactions results in a time-dependent Liouvillian for the quantum master equation. Since we are dealing with an open quantum system, we included the effect of spin relaxation and spin-selective recombination of states as well. 

Time-resolved transient absorption (TA) spectroscopy is a valuable technique for analyzing the magnetic field effects (MFEs). The MFE time profile, known as $\Delta \Delta A$, shows the difference between TA signals in the absence and the presence of the magnetic field \cite{ref:maeda2002}. We used our quantum-based model to calculate $\Delta \Delta A$ signal theoretically (with some reasonable simplifying assumptions regarding FAD photochemistry) as a way to validate the presented model. In contrast to previous semi-classical models, the quantum-based model is capable of predicting the full magnetic field dependence, and all interactions are treated systematically by incorporating through the Liouvillian of the quantum master equation. Our results are consistent with the experimental observations.

This paper is organized as follows. The theoretical model we used to demonstrate the magnetic field effect on the spin system based on quantum master equation is provided in Sec. \ref{sec:theory}. An overview of molecular dynamics simulation for FAD in water solvent and how this simulation can inform us about the distance between radicals within FAD molecule plus the method we used to solve the time-dependent quantum master equation is presented in Sec. \ref{sec:MD}. In Sec. \ref{sec:photochem}, we discussed the photochemistry of FAD and demonstrated a good correspondence between our theoretical model and experimental results. 

\section{Theoretical model}\label{sec:theory}
The biradical system within the FAD molecule consists of two radicals, A and B, one in the flavin and the other in adenine moiety, as shown in Fig. \ref{fig:FAD}.

Several interactions can influence the dynamics of the spin system, such as Zeeman, hyperfine, exchange, electron dipole-dipole, spin-orbit, nuclear Zeeman, and nuclear dipole-dipole couplings. However, some of these interactions are negligible for FAD biradical. The spin-orbit interaction, which arises from the coupling between electron spin and the magnetic field generated by the orbital motion of the electron, can be ignored for organic radicals with low symmetry and no heavy atomic nuclei \cite{ref:Hore, ref:spin-orbit}. 
Nuclear Zeeman coupling and dipole-dipole interaction between nuclei are also negligible compared to their electronic counterpart because the gyromagnetic ratio for nuclei is much smaller than for electrons. We introduce the four remaining couplings in the following.

The Zeeman interaction with the Hamiltonian of form
\begin{equation} \label{eq:zeeman}
    H_Z = -\gamma_e \Vec{B}.\Vec{S}
\end{equation}
is responsible for the effect of the external magnetic field on the electronic spin system, where $\gamma_e$ and $\Vec{B}$ are the gyromagnetic ratio of the electron and the external magnetic field, respectively. Also, $\Vec{S} = (\hat{S}_x, \hat{S}_y, \hat{S}_z)$ is the electron spin momentum. The Zeeman interaction splits the energy levels of a particle with spin (like electrons) under the influence of an external magnetic field. In our modeling, we assumed that the direction of the external magnetic field is aligned with the z-axis of the coordinate system. 

The hyperfine interaction takes care of the interaction between atomic nuclei and the spin of electrons. Fundamentally speaking, hyperfine coupling is composed of two interactions. The first is because of the dipole-dipole coupling between magnetic moments of electron and nucleus, analogous to the classical dipolar coupling of two magnetic moments. The other is the Fermi contact term, which is due to the non-zero probability density of the electron at the nucleus. Fermi contact is an isotropic interaction and happens in radicals with p, d, or f orbitals \cite{ref:Alvarez}. The hyperfine coupling has the following form:
\begin{equation}\label{eq:hf}
    H_{HF} = \gamma_e \Vec{S}\cdot\Vec{\Vec{A}}\cdot \Vec{I}.
\end{equation}
The hyperfine tensor $\Vec{\Vec{A}}$ can be calculated using density functional theory (DFT). Because of rotational averaging due to molecular motion in the solution we only consider isotropic hyperfine interaction. We assumed hyperfine interaction with only one spin-1/2 nucleus with a hyperfine coupling constant (HFCC) of $a=0.4$ mT for each radical, one for flavin moiety and the other for adenine moiety. We choose these values for HFCC as they provide the best agreement between our theoretical results on the photochemistry of FAD and their experimental counterparts. The effect of different HFCC values on the photochemistry of FAD is shown in the Supplementary document (Supplementary Fig. 5). Given the interaction between electron and nucleus spins via hyperfine coupling, the required Hilbert space can be defined as $S_A \otimes I_A \otimes S_B \otimes I_B$, where $S_A$ and $S_B$ represent the electronic spin or each radical, and similarly, $I_A$ and $I_B$ represent nuclei spins.

The coupling between two unpaired electrons due to the overlap of their spatial wave function is called exchange interaction:
\begin{equation}\label{eq:exchange}
    H_J = -J_0 e^{-\beta r} (\hat{S}^2 - \hat{1}),
\end{equation}
where for the biradical of FAD, $J_0 = 2.3\times 10^8\ \rm{mT}, \beta = 21.4\ \rm{nm}^{-1}$ \cite{ref:J0} and $\hat{S}^2 = \vec{S}_{A}\cdot \vec{S}_{B} $. Notice that this coupling is very large for short distances, but drops rapidly with distance. Exchange interaction stems from the exchange symmetry of electrons (indistinguishable particles and fermions) and Coulomb force in between them.\\
Similar to the interaction between nucleus and electron spins, electronic spins of two radicals are interacting as well \cite{ref:Babcock}. This interaction known as dipole-dipole interaction has the following form:
\begin{equation}\label{eq:dd}
    H_{DD} = -\frac{|\gamma_e|}{3}\frac{2.78\ \rm{mT}}{r^2} [3(\Vec{S}_{A}\cdot \Vec{r})(\Vec{S}_{B}\cdot \Vec{r}) - r^2 (\Vec{S}_{A}\cdot \Vec{S}_{B})],
\end{equation}
where $\Vec{r}$ (in nm) is the distance between two unpaired electrons. If we assume that the axis of vector $\Vec{r}$ is aligned with the z-axis of the applied magnetic field, this interaction obtains the following form, which we utilized in our simulations:
\begin{equation*}
    H_{DD} = -\frac{|\gamma_e|}{3}\frac{2.78\ \rm{mT}}{(r/\rm{nm})^3} [2\hat{S}_{A, z}\hat{S}_{B, z} - \hat{S}_{A, x}\hat{S}_{B, x} - \hat{S}_{A, y}\hat{S}_{B, y}].
\end{equation*}
Since we are dealing with a biradical moving and rotating in the solvent, the direction of $\vec{r}$ will change and therefore, the above equation is an approximation. However, as we show in the supplementary information (Supplementary Fig. 4), our overall conclusions remain largely unchanged whether we include the dipole-dipole interaction or not.

\begin{figure}[H]
    \centering
    \includegraphics[width=0.45\textwidth]{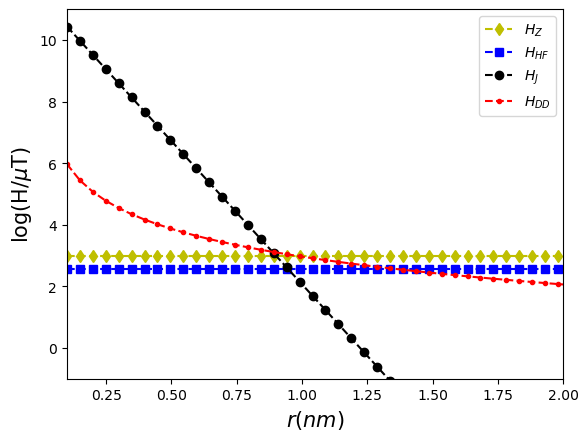}
    \caption{Logarithmic plot of the magnitudes of the Zeeman ($H_Z$), hyperfine ($H_{HF}$), exchange ($H_J$), and dipole-dipole ($H_{DD}$) interactions. The magnitude of the magnetic field is 1 mT.} 
    \label{fig:interactions_r}
\end{figure}

Exchange and dipole-dipole interactions have a dependency on the distance between two radicals meaning that it is important to be aware of the conformation of the molecule in time and the relative position of flavin and adenine moieties. Fig. \ref{fig:interactions_r} illustrates a comparison between these four abovementioned interactions for different distances. 

As evident, exchange and dipole-dipole interactions are dominant at small distances, and therefore, the effect of the Zeeman coupling is negligible. However, by gradually increasing the distance, the energy related to exchange and dipole-dipole interactions becomes negligible, and magnetic field effect can be retrieved.

Now that we understand the internal interactions of the spin system, we can model this open quantum system using quantum master equation approach to formulate the evolution of the density matrix of the spin states, $\hat{\rho}(t)$, in time. The density operator fully describes the system dynamics and demonstrates both the probabilities of each state (diagonal elements of density matrix operator) as well as coherences (off-diagonal elements of density matrix operator). 

Investigating the dynamics of this open quantum system can be done using the stochastic Liouville master equation \cite{ref:Hore}:
\begin{equation}\label{eq:mastereq}
    \frac{d}{dt} \hat{\rho}(t) = -\hat{\hat{\mathcal{L}}}(t) [\hat{\rho}(t)],
\end{equation}
where for the Liouvillian we have 
\begin{equation*}
\hat{\hat{\mathcal{L}}} = i\hat{\hat{\mathcal{H}}} + \hat{\hat{\mathcal{K}}} + \hat{\hat{\mathcal{R}}},
\end{equation*}
where $i$ is the imaginary unit and $\hat{\hat{\mathcal{H}}}, \hat{\hat{\mathcal{K}}}, \rm{and}\   \hat{\hat{\mathcal{R}}}$ are the commutator superoperator corresponding to Hamiltonian, chemical reactions, and spin relaxation, respectively. The Hamiltonian includes four interactions we have discussed above
\begin{equation*}
    \hat{{H}} = \hat{{H}}_Z + \hat{{H}}_{HF} + \hat{{H}}_{DD} + \hat{{H}}_J,
\end{equation*}
and for the Hamiltonian superoperator we have
\begin{equation*}
    \hat{\hat{\mathcal{H}}} = \hat{{H}} \otimes \hat{1} - \hat{1} \otimes \hat{{H}}.
\end{equation*}
The spin states of a radical pair can undergo different chemical reactions and we use Haberkorn model for spin-selective first-order reactions \cite{ref:Hore} with the following superoperator.
\begin{equation*}
    \hat{\hat{\mathcal{K}}} = \frac{k_s}{2}(\hat{P}^S \otimes \hat{1} + \hat{1}\otimes \hat{P}^S) + \frac{k_t}{2}(\hat{P}^T \otimes \hat{1} + \hat{1}\otimes \hat{P}^T),
\end{equation*}
where $k_s, k_t$ are the singlet and triplet reaction rates, respectively. Also, $\hat{P}^S$ and $\hat{P}^T$ are the singlet and triplet projection operators, respectively. The singlet and triplet states are defined as:
\begin{align*}
    \ket{S} &= \frac{1}{\sqrt{2}}\big(\ket{\uparrow_A\downarrow_B} - \ket{\downarrow_A\uparrow_B}\big)\\
    \ket{T_+} &= \ket{\uparrow_A\uparrow_B}\\
    \ket{T_0} &= \frac{1}{\sqrt{2}}\big(\ket{\uparrow_A\downarrow_B} + \ket{\downarrow_A\uparrow_B}\big)\\
    \ket{T_-} &= \ket{\downarrow_A\downarrow_B}.
\end{align*}
Moreover, the relaxation superoperator \cite{ref:Hore}

\begin{align*}
    \hat{\hat{\mathcal{R}}} =&k_r^A(\frac{3}{4}\hat{1}\otimes\hat{1} - \hat{S}_{A, x} \otimes \hat{S}_{A, x}^{T} - \hat{S}_{A, y} \otimes \hat{S}_{A, y}^{T} - \hat{S}_{A, z} \otimes \hat{S}_{A, z}^{T})\\+& k_r^B(\frac{3}{4}\hat{1}\otimes\hat{1} - \hat{S}_{B, x} \otimes \hat{S}_{B, x}^{T} - \hat{S}_{B, y} \otimes \hat{S}_{B, y}^{T} - \hat{S}_{B, z} \otimes \hat{S}_{B, z}^{T}),
\end{align*}
takes care of the relaxation process of the elements of the spin density matrix $\hat{\rho}(t)$. The relaxation superoperator introduced here accounts for random time-dependent local fields and spin rotation. For a time-independent system, the solution to Eq. \ref{eq:mastereq} is given by the following expression
\begin{equation*}
    \hat{\rho}(t) = e^{-\hat{\hat{\mathcal{L}}} t} [\hat{\rho}(0)],
\end{equation*}
where $\hat{\rho}(0)$ is the initial state.

\begin{figure}[H]
    \centering \includegraphics[width=0.49\textwidth]{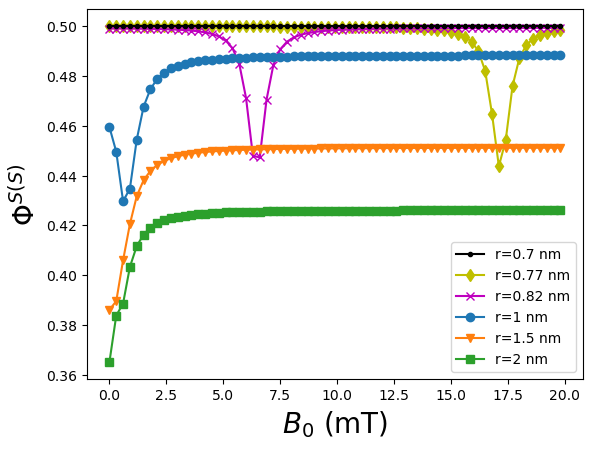}
    \caption{Singlet yield of a radical pair system as a function of the magnetic field with initial singlet state, $\Phi^{S(S)}$, for different distances between radicals. Recombination ($k_s, k_r$) and relaxation rates ($k_r^A, k_r^B$) are all set to $10^6$ $\rm{s}^{-1}$. The plot corresponding to $r=0.7$ nm is flat and equal to 0.5.}
    \label{fig:sweepB0_2}
\end{figure}

The fractional singlet yield produced by the RPM can be calculated as follows
\begin{equation*}
    \Phi^{S} = k_s\int_{0}^{\infty}\rm{Tr}\big[\hat{P}^S \hat{\rho}(t)\big] dt,
\end{equation*}
Analyzing the effect of the distance between two radicals on the singlet yield of this spin system is a key step in our calculations. The singlet yield of the system is illustrated in Fig. \ref{fig:sweepB0_2} as a function of the applied magnetic field for various values of $r$. 

As evident, the profile of the singlet yield is different for different distances due to the dependency of exchange and dipole-dipole interactions with the distance between radicals. At short distances ($<0.7$ nm), the magnitude of exchange and dipole-dipole couplings is much greater than Zeeman interaction, and therefore, no MFE can be seen (no dependency of $\Phi^{S(S)}$ on changing magnetic field). Increasing distance up to $r=1$ nm results in a dip in the plot, because exchange and dipole-dipole interactions cancel out each other at certain conditions, known as J/D cancellation \cite{ref:efimova}. At higher distances ($>1.5$ nm) we only see the effect of Zeeman and hyperfine interactions. 
A mesh-grid style graph of $\Phi^{S(S)}$ for different distances and magnetic fields is provided in Supplementary Fig. 1.

\section{Molecular dynamics simulation and time dependent quantum system}\label{sec:MD}

\begin{figure}[H]
  \subcaptionbox*{}[.49\linewidth]{
    \includegraphics[width=\linewidth]{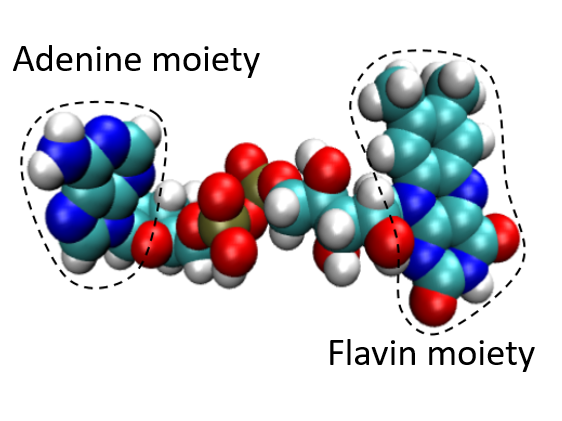}}
  \hfill
  \subcaptionbox*{}[.49\linewidth]{%
    \includegraphics[width=\linewidth]{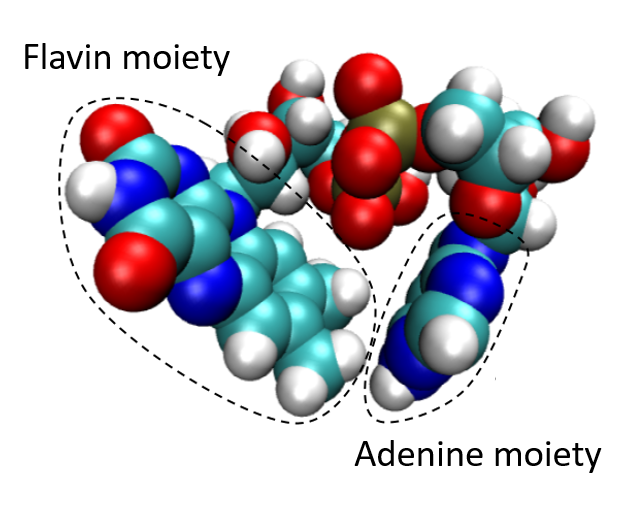}%
  }
  \caption{FAD molecule depicted in the open (left) and closed (right) configurations. The configurations are generated by GROMACS \cite{ref:GROMACS} and depicted by VMD computer program \cite{ref:VMD}.}
  \label{fig:FADconf}
\end{figure}

The fluorescence properties of FAD show that this molecule exists in closed and open conformations \cite{ref:FADfluroscence} as depicted in Fig. \ref{fig:FADconf}. Closed conformation refers to the case in which the adenine and flavin moieties stack to each other, and on the other hand, in the open conformation, the distance between these two moieties is larger than the former case \cite{ref:open/close}. 

Molecular dynamics simulation enables us to simulate and predict the relative position and movement of every atom in a molecular system for a given time. The FAD molecule samples a wide range of intermediate states as it fluctuates between open and closed conformations in a water solvent \cite{ref:MD}. By performing MD simulation, one can obtain the conformation of the FAD molecule in time.

The MD simulation was performed using GROMACS with CHARMM36 force field. We used Protein Data Bank file 6PTZ \cite{ref:pdb} to extract the initial structure of FAD and hydrogen atoms were added to the molecular structure using Avogadro. The FAD molecule was placed in a cubic unit cell with at least 1 nm distance from the edges and surrounded by 1478 water molecules (single point-charge water model). The charged solvated system has been neutralized by adding ions and afterward we relaxed the structure by performing energy minimization. Before running the MD simulation, we equilibrated the system under two conditions, first, constant number of particles, volume, and temperature, and second, constant number of particles, pressure, and temperature ensembles. While the former stabilizes the temperature of the system, the latter stabilizes pressure. Following energy minimization and equilibration processes, MD production was performed for a duration of $1\ \mu$s. We chose this simulation time to make sure that all conformation states have been sampled sufficiently. The time step for the MD simulation was 2 fs. In order to capture the stochastic behavior of this molecular system, we performed four MD simulations with the same conditions.

As an estimate of the distance between two radicals, we calculated the distance between the center of mass (COM) of the middle ring in the flavin moiety and COM of the 6-membered ring of the adenine moiety. We have chosen these two points as proxies of the unpaired electrons in the molecular orbitals of molecule FAD.

Previous experimental studies showed the effect of different pH values of the solvent on the magnetic field effect of FAD \cite{ref:woodward, ref:maeda2005}. Different pH values change the protonation state of the molecule, and thus, lead to different conformation spaces for the dynamics of FAD in water solvent. These changes result in a different time profile for the distance between COMs, and subsequently, different magnetic field effects. The MD simulation performed for this study only considers the neutral case of pH value 7 which is close to biologically relevant pH values.

Performing MD simulation results in many data points for the distance between radicals which need to be fed into the quantum master equation. To solve the quantum master equation with time-dependent Liouvillian, we assumed that Liouvillian is constant during very small time steps, and calculated its evolution in time by solving the following recurring equation
\begin{equation*}
    \hat{\rho}_{i+1} = e^{-\hat{\hat{\mathcal{L}}}_{i} \Delta t} \hat{\rho}_{i},
\end{equation*}
where $\Delta t$ is the time steps,  $\hat{\rho}_{i+1}$ is the density operator at time $(i+1)\Delta t$, and $\hat{\hat{\mathcal{L}}}_{i}, \hat{\rho}_{i}$ are the Liouvillian and density operator at time $i\Delta t$, respectively. Notice that $\hat{\rho}_0 = \hat{\rho}(0)$. In our study, all calculations related to solving the master equation were performed using the Python programming language.

In order to avoid the extreme computational cost in solving the master equation for many different distances, we averaged them over time windows of length 1 ns. We calculated $\hat{\rho}_{i+1}$ for each time step of length $\Delta t = 1$ ns using Python. For this purpose, we used an iterative algorithm. We first constructed the matrices for the Hamiltonian, chemical reaction, and spin relaxation superoperators at some given time step $i\Delta t$, and used them to calculate the corresponding matrix for the Liouvillian superoperator also at $t = i\Delta t$. Using this Liouvillian superoperator matrix we computed the matrix corresponding to $e^{-\hat{\hat{\mathcal{L}}}_{i}\Delta t}$ and multiplied it to $\hat{\rho}_i$ matrix to get the density matrix at the next time step, i.e., $\hat{\rho}_{i+1}$. We then repeated this process the required number of times.

\begin{figure}[H]
    \centering
    \includegraphics[width=0.47\textwidth]{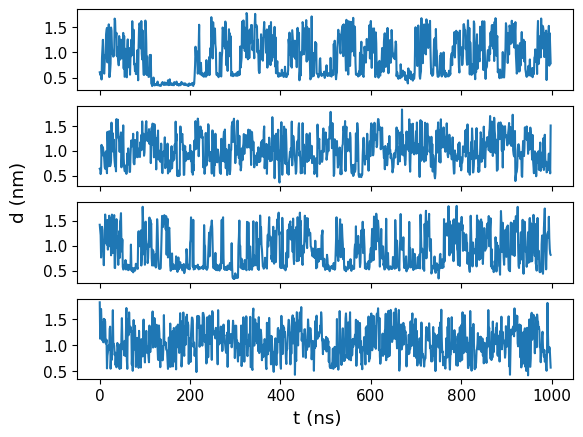}
    \vspace{-0.2cm}
    \caption{Distance between centers of mass of the adenine and isoalloxazine ring. Four molecular dynamics simulations for FAD in water solvent have been performed to capture the stochastic behavior. These plots are the averaged version of results from MD simulation with 1000 time windows of length each 1 ns.}
    \label{fig:sampledMD}
\end{figure}

The results on the distance between the COMs are depicted in Fig. \ref{fig:sampledMD}. These plots are the averaged version of the original results obtained from MD simulation. See Supplementary for the original plots before averaging (Supplementary Fig. 2).

\section{Magnetic field effects on the transient absorption}\label{sec:photochem}
\begin{figure}[H]
    \centering
    \includegraphics[width=0.45\textwidth]{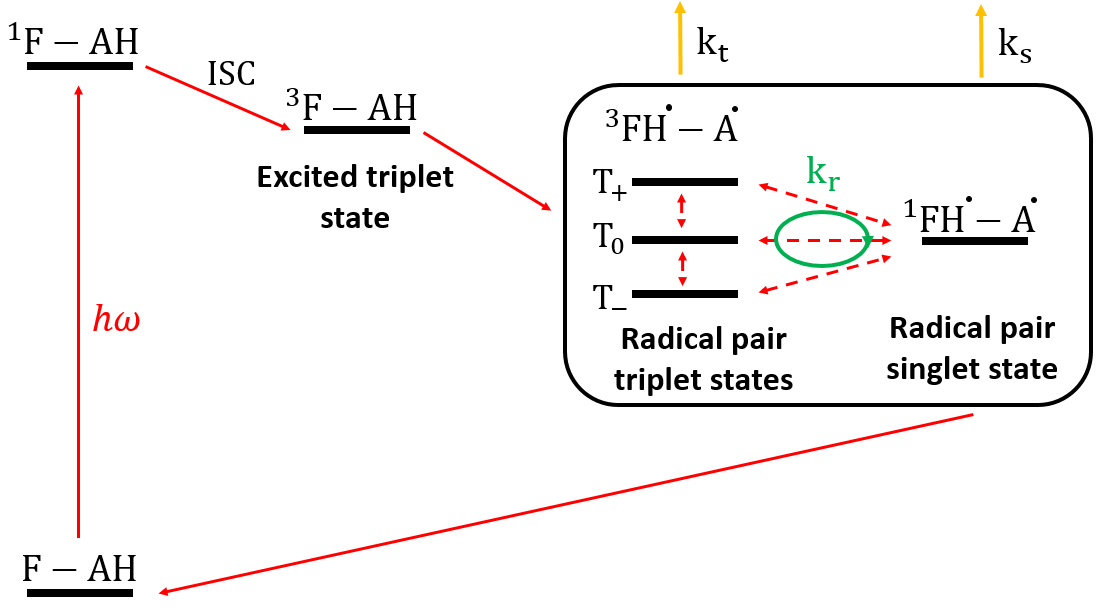}
    \caption{Reaction scheme for photochemistry of FAD. $k_s, k_t$ are denoting recombination rates for singlet and triplet states, respectively. The relaxation rate, $k_r$, is equal for both singlet and triplet states.}
    \label{fig:FADphotochemistry}
\end{figure}

In this section, we make a connection between our quantum theoretical model and existing experimental observations. To do so, a brief understanding of the FAD photochemistry is required. Under blue light excitation of the flavin moiety and an intramolecular electron transfer from the adenine moiety, a biradical can be formed. This biradical undergoes spin-selective recombination resulting in magnetic field effects on the photochemistry of FAD \cite{ref:woodward}. The scheme showing the FAD photochemistry is provided in Fig. \ref{fig:FADphotochemistry}. 

Several chemical kinetic models have been proposed to explain the magnetic field effects on the photochemistry of FAD at different pH values \cite{ref:maeda2005, ref:kaptein}. However, these models are semi-classical and do not involve a full quantum mechanical treatment of radical pair dynamics. This restricts their predictive power to the two extreme cases of low and high magnetic fields only.

In the quantum approach, the state of the radical pair is described by the density matrix. The density matrix in the basis of singlet-triplet states ($S, T_+, T_0, $ and $T_-$) can be described as follows:
\begin{equation*}
    \hat{\rho}(t) = \begin{pmatrix}
        \rho_{SS} & \rho_{T_+ S} & \rho_{T_0 S} & \rho_{T_- S}\\
        \rho_{S T_+} & \rho_{T_+ T_+} & \rho_{T_0 T_+} & \rho_{T_- T_+}\\
        \rho_{S T_0} & \rho_{T_+ T_0} & \rho_{T_0 T_0} & \rho_{T_- T_0}\\
        \rho_{S T_-} & \rho_{T_+ T_-} & \rho_{T_0 T_-} & \rho_{T_+ T_-}
    \end{pmatrix}(t).
\end{equation*}
The diagonal and off-diagonal elements of the density matrix show the concentration of each state and the coherence, respectively. 
As shown in an earlier section, the time evolution (Eq. \ref{eq:mastereq}) of the density matrix is governed by the master equation. This approach allows a systematic treatment for all magnetic field values and can also incorporate the effects of varying distances between the radicals. 

On the other hand, in the semi-classical approach, the state of the radical pair is described by a population vector whose time evolution is governed by a rate equation. Rate equations are a set of equations relating the rate of change of an individual state population (its first derivative in time) to a linear combination of the population of all states. The semi-classical approach only involves the diagonal elements of the density matrix in the S-T basis (the populations) and neglects the coherences. We found that neglecting the coherences turns out to be a fair assumption for this radical pair system as they die out very quickly (see supplementary Fig. 3).
The form of the rate equations for this model is given as follows.  
\begin{equation*}
    \frac{d}{dt} \begin{pmatrix}
        [S] \\ 
        [T_+] \\
        [T_0] \\
        [T_-]
    \end{pmatrix} = \begin{pmatrix}
        k_{SS} & \hdots & k_{ST_-}\\
        \vdots & \ddots & \vdots \\
        k_{T_-S} & \hdots & k_{T_-T_-}
    \end{pmatrix}\begin{pmatrix}
        [S] \\ 
        [T_+] \\
        [T_0] \\
        [T_-]
    \end{pmatrix},
\end{equation*}
where $[S], [T_+], [T_0], $ and $[T_-]$ are the population of different states and $k_{ij}$ are the rates for transition between states. Notice that the magnitude of different interactions defines the rates. As discussed in \cite{ref:maeda2005}, in the absence of magnetic field (B=0), only the hyperfine coupling with the rate of $k_{\rm{hfc}}$ is responsible for the interconversion between these four states. $k_{\rm{hfc}}$ depends on the effective local magnetic field of each nucleus and has the following expression
\vspace{-0.4cm}
\begin{equation*}
    k_{\rm{hfc}} = \frac{2(B_1^2 + B_2^2)}{\hbar (B_1 + B_2)}.
\end{equation*}
In order to calculate $B_1$ and $B_2$, we have
\vspace{-0.2cm}
\begin{equation*}
    B_i(i=1, 2) = \Bigg(\sum_k \Big(I_{ik}(I_{ik}+1)a_{ik}^2\Big)\Bigg)^{\frac{1}{2}},
\end{equation*}
where $a_{ik}$ and $I_{ik}$ represent the HFCC and nuclear spin quantum number of the $k$th nucleus in the $i$th radical, respectively. On the other hand, in the presence of a large magnetic field, the energy level of $T_+$ and $T_-$ states are separated from $S$ and $T_0$ states due to the Zeeman energy. As explained in \cite{ref:maeda2005}, this implies that the electron spin relaxation rate ($k_r$) governs the interconversion between $T_+$ and $T_-$ states and $S, T_0$ states, and $k_{\rm{hfc}}$ is responsible for the interconversion between $S$ and $T_0$ states.

As it is evident, this model fails to predict the behavior of system for an intermediate magnetic field, where hyperfine and Zeeman interactions are not negligible to each other. Moreover, this model ignores the effect of other important interactions such as exchange and dipole-dipole couplings. As we saw previously, no magnetic field effect can be observed at small distances between two radicals. Without including the effect of the distance, the model cannot cannot fully capture the quantum dynamics.

\begin{figure}[H]
  \subcaptionbox*{(a) B = 0}[.48\linewidth]{
    \includegraphics[width=\linewidth]{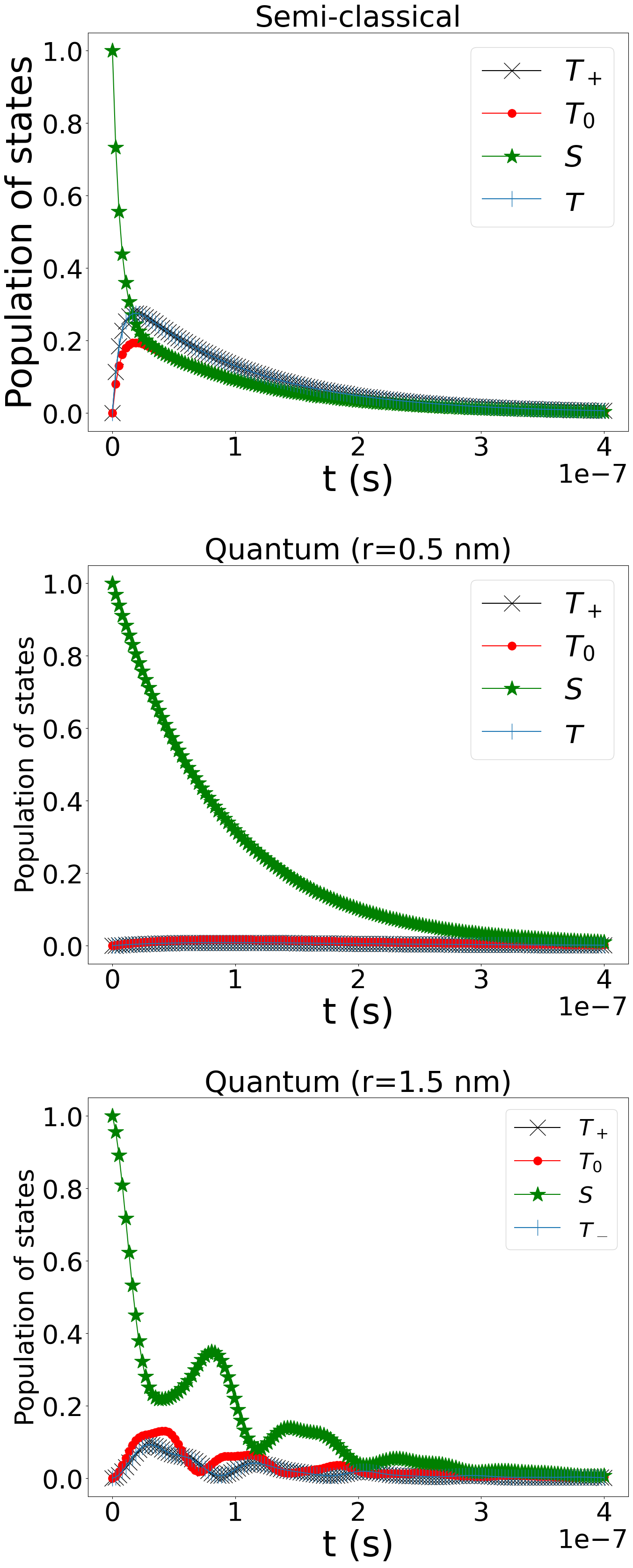}}
  \hfill
  \subcaptionbox*{(b) B = 20 mT}[.48\linewidth]{%
    \includegraphics[width=\linewidth]{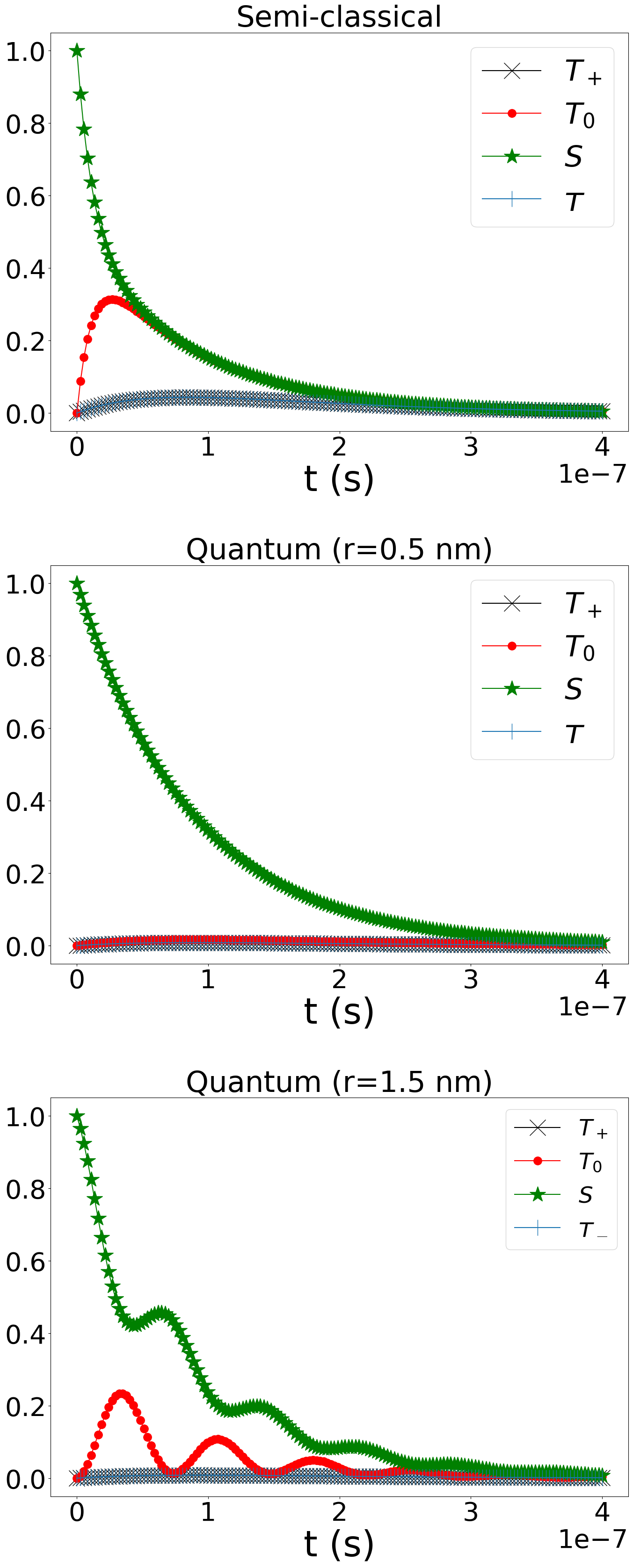}%
  }
  \caption{Time evolution of the singlet ($S$) and triplet ($T_+, T_0$, and $T_-$) states for two different cases: (a) B=0 (low magnetic field) and (b) B=20 mT (high magnetic field). The initial state is the singlet state. Relaxation ($k_r^A, k_r^B$) and recombination rates ($k_s, k_t$) are $10^6$ and $10^7\ \rm{s}^{-1}$, respectively.}
  \label{fig:populations}
\end{figure}

As a comparison between quantum master equation and semi-classical rate equation approaches, let us look at the population of singlet and triplet states for these two models under the same conditions. Results are illustrated in Fig. \ref{fig:populations}.

In Fig. \ref{fig:populations}, notice that in the middle plot ($r=0.5$nm), no interconversion between singlet and triplet states can be observed resulting in no magnetic field effect. This is due to strong exchange and dipole-dipole interactions at small distances. Also, comparing the right and left graphs shows some oscillations for the quantum model. This indicates that semi-classical model cannot fully capture the quantum dynamics of the system correctly.

Previously, semi-classical models have mainly been used to explain the changes in the transient absorption (TA) signal in the presence and absence of a magnetic field. The idea behind TA spectra is to first excite the sample with a pump pulse and then after a period of time, send a probe pulse to obtain the absorption spectra of the excited sample. 

The time profile of MFE is calculated by subtracting TA signals for zero and high ($B = B_0$) magnetic fields. 
\vspace{-0.1cm}
\begin{equation}
    \Delta \Delta A_{(B = B_0, t)} = \Delta A_{(B = B_0, t)} - \Delta A_{(B = 0, t)}
    \label{eq:dda}.
\end{equation}
As treated by Murakami et al. \cite{ref:maeda2005}, the time-resolved MFE action signal is given by $\Delta \Delta A(t) = \epsilon_R \Delta C_R(t) + \epsilon_T \Delta C_T(t)$, where $\Delta C_R, \Delta C_T$ represent the contribution of radical pair triplet state and excited triplet state (see Fig. \ref{fig:FADphotochemistry}) on the MFE action spectra, respectively. $\epsilon_R, \epsilon_T$ are the template spectra for their corresponding states.

\begin{figure}[H]
    \centering \includegraphics[width=0.4\textwidth]{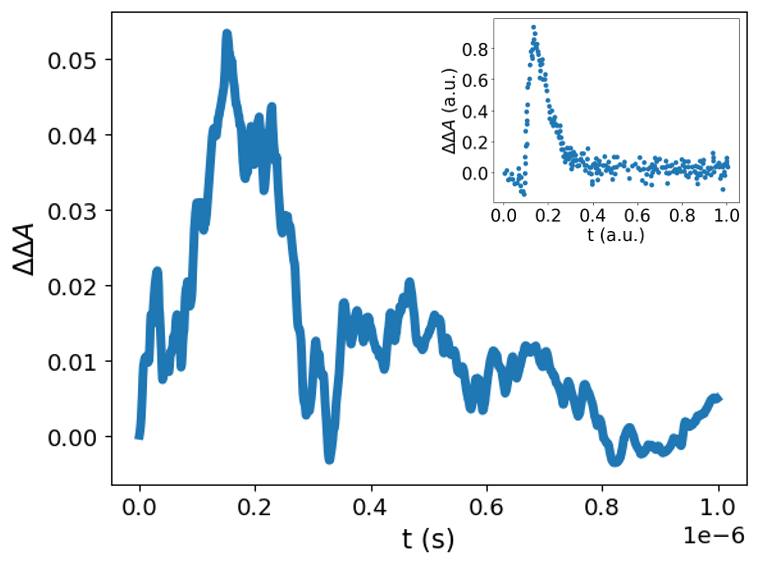}
    \vspace{-0.3cm}
    \caption{Time profile of transient absorption $\Delta \Delta A$. Both relaxation ($k_r^A, k_r^B$) and recombination ($k_s, k_t$) rates are $10^6\ \rm{s}^{-1}$ for the main curve. The inset displays a typical experimental $\Delta \Delta A$ time profile (pH=3.3) and was extracted manually from \cite{ref:maeda2005}. The experimental signal represents an averaging over a large number of FAD molecules. The Magnetic field is 0.2 T for both main and inset curves. }
    \label{fig:dda_time}
\end{figure}

However, the photochemistry model proposed by Murakami et al. \cite{ref:maeda2005} is valid for low pH values. In their model, there is electron transfer/back electron transfer between excited triplet state and radical pair triplet state. In higher pH values, there is only the feeding term from the excited triplet state to the radical pair triplet state (The feeding term from the radical pair triplet state to the excited triplet states decays as the concentration of protons ($[H^+]$) increases) \cite{ref:woodward}. Under this condition, starting from the radical pair triplet state works well for analyzing the dynamics of the system. Thus, for simplification, we only take into account the contribution of radical pair triplet state while calculating $\Delta \Delta A$, i.e., we assumed $\Delta C_T = 0$.
Implementing the population of the triplet state, up to a scaling factor, into Eq. \ref{eq:dda} \cite{ref:maeda2005} results in
\begin{align*}
    \Delta \Delta A = \rm{fac}\Bigg(&\big([T_+] + [T_0] + [T_-]\big)_{B = B_0}\\ - &\big([T_+] + [T_0] + [T_-]\big)_{B = 0}\Bigg).
\end{align*}
Notice that in the quantum model, the concentrations are defined as $[T_i] = \rho_{ii}$. The calculated $\Delta \Delta A$ signal obtained from the quantum master equation is illustrated in Fig. \ref{fig:dda_time}. The inset curve in this figure shows a typical experimental $\Delta \Delta A$ signal for pH 3.3. Although our calculation contains several assumptions, our theoretical curve follows the trend in a typical experimental plot and is in qualitative agreement with them. The experimental curve corresponds to averaging over many individual FAD molecules and includes the contribution from excited triplet states as well.
\vspace{-0.1cm}
Another related quantity that is used to show that the photochemistry of FAD is sensitive to magnetic field is MARY (Magnetically Affected Reaction Yield) curves \cite{ref:woodward, ref:mary}. The MARY curve can be obtained by dividing $\Delta \Delta A_{(B=B_0, t)}$ by $\Delta A_{(B=0, t)}$ and integrating over the whole time window \cite{ref:woodward}. 
\begin{equation}\label{eq:MARY}
    \rm{MFE} \% = \int_{0}^{\infty} \frac{\Delta \Delta A_{(B = B_0, t)}}{\Delta A_{(B=0, t)}} dt \times 100\%.
\end{equation}

\begin{figure}[H]
    \centering
    \includegraphics[width=0.39\textwidth]{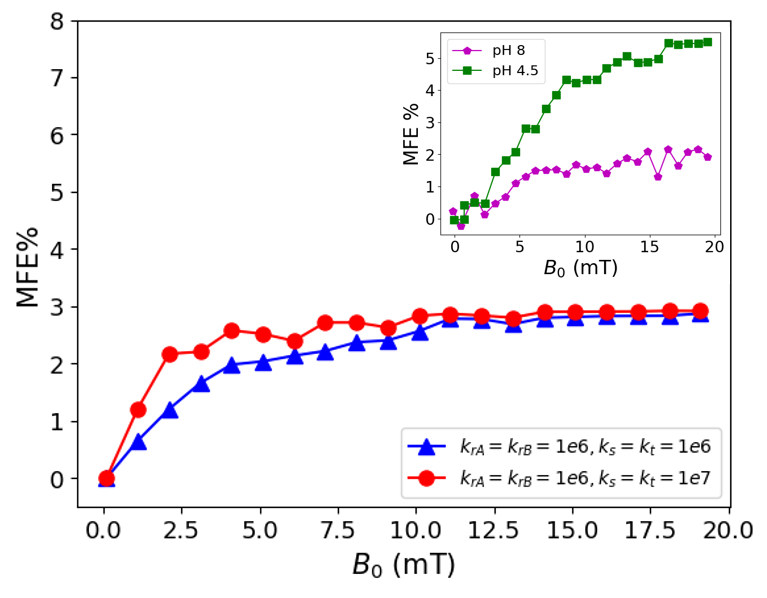}
    \vspace{-0.2cm}
    \caption{Calculated magnetic field effects (MFEs) of FAD (based on our quantum theory model) for different relaxation and reaction rates. This plot shows the magnetically affected reaction yield (MARY) spectra as defined in Eq. \ref{eq:MARY}. The main plots are valid for pH 7. The inset displays experimental MFEs for different pH values. Data points extracted manually from Fig. 3(a) in \cite{ref:woodward} for inset.}
    \label{fig:MFE}
\end{figure}

The calculated MARY curves are illustrated in Fig. \ref{fig:MFE} for different relaxation and recombination rates. The inset in this figure shows experimental MARY spectra for different pH values. 
The results in Fig. \ref{fig:MFE} are similar to the experimental results obtained by Antill et al. (illustrated in the inset of Fig. \ref{fig:MFE}) \cite{ref:woodward}. A similar trend can be observed. This verifies that the proposed RPM model is able to explain the FAD photochemistry is magnetic field sensitive. It indicates the effect of an external magnetic field on the interconversion between singlet/triplet states, which results in a nonzero MARY spectra at high enough magnetic fields. 

\section{Discussion}
In the present work, we investigated the MFEs on the spin dynamics of a biradical formed within molecule FAD, which may be a possible biological magnetic field sensor. We used RPM model to evaluate how magnetic field changes spin dynamics and photochemistry of FAD. 

The population of singlet and triplet states is obtained by solving Lindbald quantum master equation for a biradical formed in an FAD molecule. Our quantum master equation includes Zeeman, hyperfine, exchange, and dipole-dipole interactions. Other interactions such as spin-orbit, nuclear Zeeman, and nuclear dipole-dipole interactions are ignored because of their weak effect in our system. The openness of this system is modeled via relaxation and recombination processes. 

Exchange and dipole-dipole interactions are dependent on the distance between two radicals and altering this distance results in a different MFE. 
To study different conformations of FAD, we performed MD simulation of molecule in a water solvent to obtain its configuration in time.  
Fig. \ref{fig:sweepB0_2} illustrates changes in singlet quantum yield for different distances and magnetic fields. MFE can be observed only for relatively open conformation of molecule. In addition, MD simulation results indicate that this molecule has non-zero lifetime in open conformation, Fig. \ref{fig:sampledMD}, and as our theoretical model predicts, this lifetime is enough to observe non-zero MFE.

Our quantum-based model which is powered by MD simulation has several advantages over previous semi-classical model. The semi-classical model formulates the spin dynamics of the biradical within FAD via rate equations and ignores the effect of various interactions. Because of this, the model is limited to only low and high magnetic field cases and cannot capture the effect of an intermediate magnetic field. The density matrix operator and its elements are the core of our calculations in this quantum model. Differences between quantum and semi-classical models are illustrated in Fig. \ref{fig:populations}, showing the singlet and triplet time evolution.

As a method to verify our model and compare it to experimental observations, we calculated the transient absorption (TA) signal, as shown in Fig. \ref{fig:dda_time} and Fig. \ref{fig:MFE}, and compared it with experimental results from TA spectroscopy. It is worth keeping in mind that our theoretical calculation of this signal contains some assumptions like ignoring the role of the excited triplet state as discussed in previous sections. The comparison shows good correspondence between our theoretical model and experimental results. 

As a suggestion for future research on this topic, there is room for performing accurate MD simulations for different conditions. In Sec. \ref{sec:MD}, we discussed briefly the effect of different pH values for the solvent surrounding FAD. In our MD simulation, parameters of the system such as bond lengths, angles, and dihedrals were taken from standard building blocks of molecule FAD. Different protonation states require different parameters and these new parameters can be obtained by a process known as parametrization which requires ab initio calculations. In addition, one can include the role of excited triplet state into calculating $\Delta \Delta A$ signal, i.e., add this state into the quantum master equation. On the experimental side, more \textit{in vivo} and \textit{in vitro} (at pH 7) experiments are required, for example on magnetic field dependence, to understand the role of FAD biradicals in magnetosensitivity in various organisms. Given our quantum-theoretical model for FAD biradicals, a better insight into such experiments will be available. Also, This quantum model based on RPM can be applied to other studies about the MFEs on the circadian clock, neurogenesis, and stem cell growth. \cite{ref:circadianHadi, ref:neuroHadi, ref:stemcellWendy}.

\section*{Acknowledgments}
This work was supported by the Natural Sciences and Engineering Research Council of Canada through its Discovery Grant Program and the Alliance Quantum Consortia Grant `Quantum Enhanced Sensing and Imaging (QuEnSI)'. The authors would like to thank Dennis Salahub, Belinda Heyne, Peter Tieleman, Hristina Zhekova, Valentina Corradi, DB Sridhar, and Rishabh Shukla for their valuable input.

\pagebreak
\widetext
\begin{center}
\textbf{\large Supplementary Materials}
\end{center}

Supplementary Fig. 1 (following Fig. 3 in the paper) This 2D plot shows $\Phi^{S(S)}$ for different magnetic fields ($B_0$) and distance ($r$). Singlet fraction yield: $\Phi^{S(S)} = k_s \int_0^\infty \rm{Tr}\big[\hat{P}^S \hat{\rho}(t)\big]dt$ (initial state: singlet state).

\begin{figure}[H]
    \centering
    \includegraphics[width=0.49\textwidth]{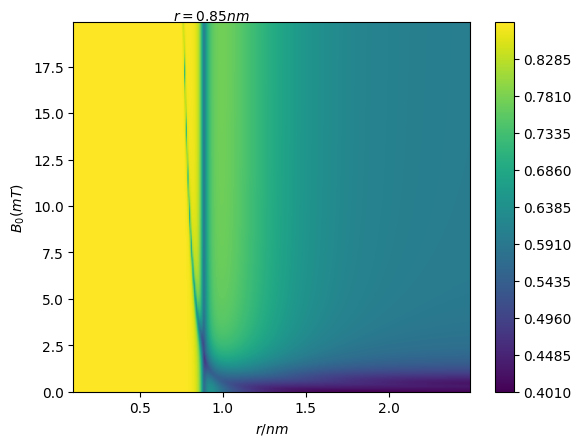}
    \caption*{\textbf{Supplementary Fig. 1}: Singlet yield of a radical pair system with initial singlet state.}
    \label{figS:1}
\end{figure}

\vspace{0.75cm}

Supplementary Fig. 2 (following Fig. 5 in the paper) The original results on the distance between centers of mass of the Adenine and isoalloxazine ring in FAD, calculated using GROMACS. The average of these graphs is illustrated in main text.

\begin{figure}[H]
    \centering
    \includegraphics[width=0.49\textwidth]{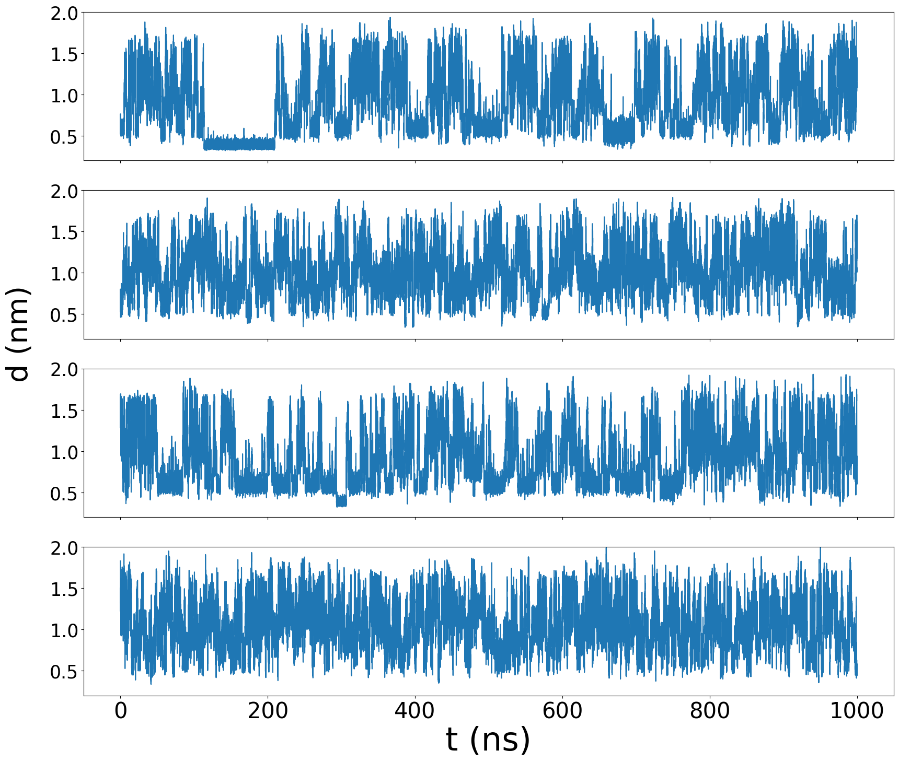}
    \caption*{\textbf{Supplementary Fig. 2}: Distance between centers of mass of the Adenine and isoalloxazine ring. Results obtained from MD}
    \label{figS:2}
\end{figure}

Supplementary Fig. 3 (following Fig. 7 in the paper) As mentioned in paper, the off-diagonal terms of density operator ($\rho(t)$) are negligible. 

\begin{figure}[H]
  \subcaptionbox*{(a) $r = 0.5$ nm}[.49\linewidth]{
    \includegraphics[width=\linewidth]{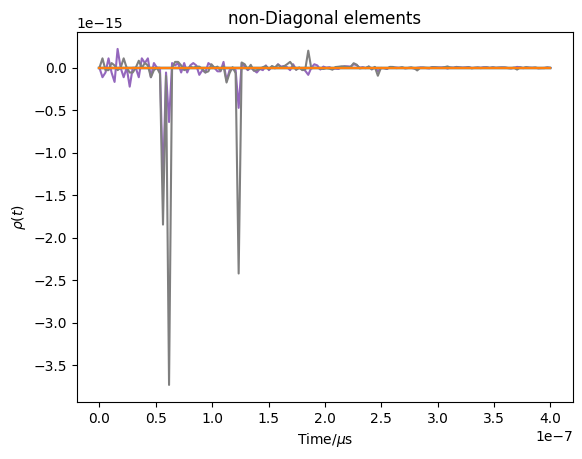}}
  \hfill
  \subcaptionbox*{(b) $r = 1.5$ nm}[.49\linewidth]{%
    \includegraphics[width=\linewidth]{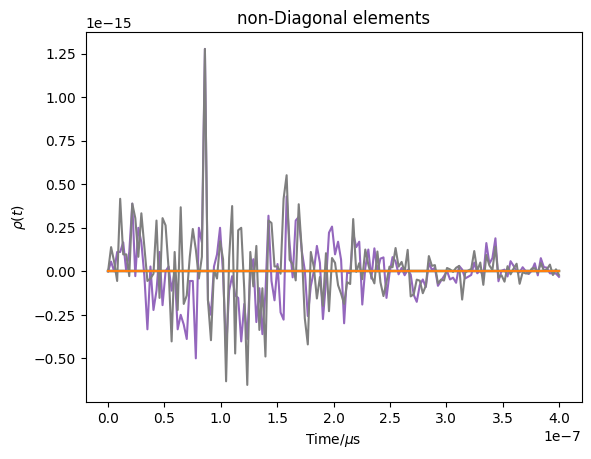}%
  }
  \caption*{\textbf{Supplementary Fig. 3}: Time evolution of the off-diagonal term of density operator for two different distances between radicals.}
  \label{figS:3}
\end{figure}

\vspace{0.75 cm}

Supplementary Fig. 4 (following Fig. 10 in the paper) As discussed in paper, ignoring dipole-dipole interaction does not change the overall conclusion about sensitivity of FAD photochemistry to magnetic field.
\begin{figure}[H]
    \centering
    \includegraphics[width=0.6\textwidth]{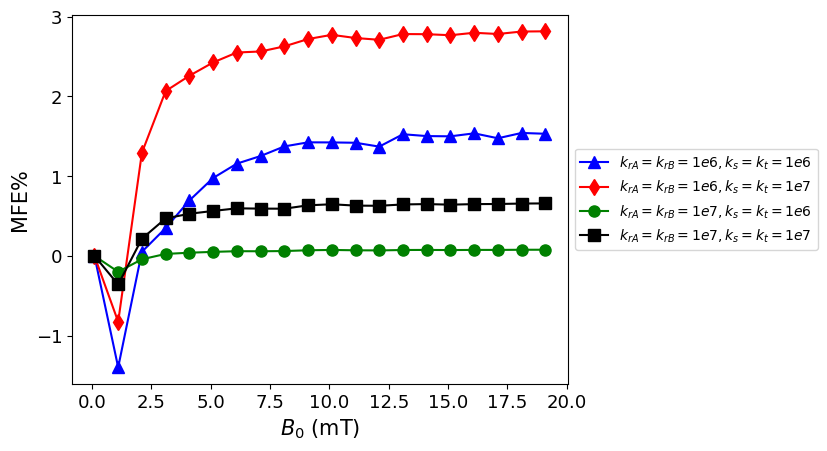}
    \caption*{\textbf{Supplementary Fig. 4}: MFEs of FAD when dipole-dipole interaction is ignored, for different relaxation and reaction rates.}
    \label{figS:4}
\end{figure}
\newpage
Supplementary Fig. 5. These curves are showing the effect of different HFCC on the MFE of FAD, including the strongest HFCC for flavin ($|a|$=0.8029) [S1].
\begin{figure}[H]
    \centering
    \includegraphics[width=0.6\textwidth]{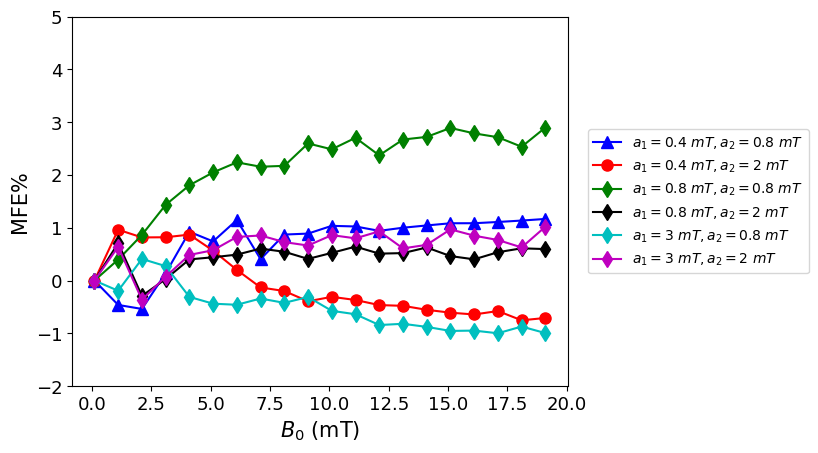}
    \caption*{\textbf{Supplementary Fig. 5}: MFEs of FAD for different HFCC of each radical. The relaxation and recombination rates are $10^6  s^{-1}$.}
    \label{figS:5}
\end{figure}

[S1] A. A. Lee, J. C. Lua, H. J. Hogben, T. Biskup, D. R. Kattnig, P. Hore, ``Alternative radical pairs for cryptochrome based magnetoreception”, \textit{Journal of the Royal Society Interface}, \textbf{11} (2014). \href{https://royalsocietypublishing.org/doi/10.1098/rsif.2013.1063}{DOI}
\end{document}